\documentclass[fleqn,usenatbib]{mnras}
\usepackage{newtxtext,newtxmath}
\usepackage[T1]{fontenc}
\DeclareRobustCommand{\VAN}[3]{#2}
\let\VANthebibliography\thebibliography
\def\thebibliography{\DeclareRobustCommand{\VAN}[3]{##3}\VANthebibliography}
\usepackage{graphicx}
\usepackage{amsmath}
\usepackage{fix-cm}

\newcommand{\frb}{FRB\,20200120E}
\newcommand{\DMunit}{pc\,cm$^{-3}$}

\title[A search for optical bursts from FRB\,20200120E]{A search for optical bursts from the repeating fast radio burst in the M81 globular cluster, FRB\,20200120E}

\author[V. S. Dhillon et al.]{V. S. Dhillon$^{1,2}$\thanks{E-mail: vik.dhillon@sheffield.ac.uk (VSD)},
L. G. Spitler$^{3}$,
K. Lackeos$^{3}$,
A. N. Mbole$^{1}$,
W. J. Page$^{1}$,
A. J. Brown$^{4}$,
N. Castro Segura$^{5}$,
\newauthor
M. J. Dyer$^{1,6}$,
J. A. Garbutt$^{1}$,
M. J. Green$^{7,8}$,
D. Jarvis$^{1}$,
M. R. Kennedy$^{9}$,
P. Kerry$^{1}$,
S. P. Littlefair$^{1}$,
\newauthor
J. McCormac$^{5}$,
J. Munday$^{10}$,
S. G. Parsons$^{1}$,
I. Pelisoli$^{5}$,
E. Pike$^{1}$,
D. I. Sahman$^{1}$,
and A. Yates$^{1}$
\\
$^{1}$Astrophysics Research Cluster, School of Mathematical and Physical Sciences, University of Sheffield, Sheffield S3 7RH, UK \\
$^{2}$Instituto de Astrof\'{i}sica de Canarias, E-38205 La Laguna, Tenerife, Spain \\
$^{3}$Max-Planck-Institut f\"{u}r Radioastronomie, Auf dem H\"{u}gel 69, D53121 Bonn, Germany \\
$^{4}$Hamburger Sternwarte, University of Hamburg, Gojenbergsweg 112, 21029 Hamburg, Germany \\
$^{5}$Astronomy and Astrophysics Group, Department of Physics, 
University of Warwick, Coventry, CV4 7AL, UK \\
$^{6}$Research Software Engineering, University of Sheffield, Sheffield, S1 4DP, UK \\
$^{7}$Homer L. Dodge Department of Physics and Astronomy, University of Oklahoma, 440 W. Brooks Street, Norman, OK 73019, USA
\\
$^{8}$JILA, University of Colorado and National Institute of Standards and Technology, 440 UCB, Boulder, CO 80309-0440, USA \\
$^{9}$School of Physics, University College Cork, Cork, T12 K8AF, Ireland \\
$^{10}$Institute for Physics and Astronomy, University of Potsdam, Karl-Liebknecht-Str. 24/25, 14476 Potsdam, Germany
}

\date{Accepted 2026 November 16. Received 2026 November 5; in original form 2026 June 24}

\pubyear{\the\year{}}
\begin{document}
\label{firstpage}
\pagerange{\pageref{firstpage}--\pageref{lastpage}}
\maketitle

\begin{abstract}
We present a search for optical bursts from the repeating fast radio burst FRB\,20200120E, which is located in a globular cluster of M81 and is the closest known extragalactic FRB. We observed FRB\,20200120E simultaneously with the high-speed, quintuple-beam optical camera HiPERCAM on the 10.4\,m Gran Telescopio Canarias and the Effelsberg 100\,m Radio Telescope. We detected one radio burst of fluence $\mathcal{F}_{\rm R} = 0.14\pm0.01$\,Jy\,ms during our observations, but found no evidence for corresponding optical bursts in the 11.98\,ms frames. We find 5$\sigma$ upper limits on the optical burst fluence of $\mathcal{F}_{\rm O} < 1.8\times10^{-3}$\,Jy\,ms (combined $u_s g_s r_s i_s z_s$-bands) and the optical burst energy of $E_{\rm O} < 5.4\times10^{36}$ erg ($r_s$-band), corrected for interstellar extinction. We compare our upper limit on the optical-to-radio fluence ratio $\mathcal{F}_{\rm O}/\mathcal{F}_{\rm R} < 0.013$ to various emission models involving inverse-Compton scattering in the magnetosphere or nebula of a highly-magnetised neutron~star.
\end{abstract}


\section{Introduction}
%original para
%Fast radio bursts (FRBs) are bright, millisecond-duration radio flashes of %unknown physical origin. First discovered in 2007 \citep{lorimer07}, there %are now over 3000 catalogued FRBs\footnote{\url{https://www.wis-tns.org}}. %The frequency-dependent arrival times of the radio bursts implies passage %through a high column density of low density plasma, which is interpreted as %the intergalactic medium, placing FRBs at cosmological distances. This makes %FRBs potentially valuable cosmological probes, providing a direct %measurement of the cosmic density of ionised baryons in the intergalactic %medium (e.g. \citealt{keane16}).

Fast radio bursts (FRBs) are bright, millisecond-duration radio flashes of unknown physical origin. First discovered in 2007 \citep{lorimer07}, there are now over 3000 catalogued FRBs\footnote{\url{https://www.wis-tns.org}}, and their extragalactic origin has been established though localising some sources to their host galaxies \citep[e.g.][]{2017Natur.541...58C, 2019Sci...365..565B}. Due to their cosmological distances, the frequency-dependent arrival times of the radio bursts are generally dominated by the low density, ionised gas in the intergalactic medium, making FRBs potentially valuable cosmological probes that provide a direct measurement of the cosmic density of ionised baryons (e.g. \citealt{2020Natur.581..391M}).

The nature of FRB progenitors is highly debated (see \citealt{platts19})\footnote{\url{http://frbtheorycat.org}}. The short durations of FRBs, with sub-structure down to tens of nanoseconds \citep[e.g.][]{nimmo21, nimmo22}, suggest that their central engines are stellar-mass compact objects such as neutron stars (e.g. \citealt{beloborodov17}, \citealt{metzger17}). Models involving highly-magnetised neutron stars, known as magnetars, received a boost due to the discovery of a bright millisecond-duration radio burst from the Galactic magnetar SGR\,1935+2154 (\citealt{bochenek20}, \citealt{chime20a}). However, the rate of SGR\,1935+2154-like bursts can not explain the observed extragalactic FRB rate without including an additional population of magnetars with stronger magnetic fields \citep{margalit20}.

%original para
%One of the difficulties in studying FRBs is that their positional %uncertainties are of order arcminutes, making follow-up observations at %other wavelengths challenging. The discovery of the first FRB to show %repeated bursts, FRB\,121102 \citep{spitler16}, finally enabled radio %interferometers to refine the position of an FRB to better than an %arcsecond. Over 80 repeaters are now known, approximately half of which have %been localised to host galaxies \citep{ibik24}, thereby confirming the %extragalactic nature of FRBs. Three of these repeaters also show tentative %evidence for periodicities in their burst arrival times (e.g. %\citealt{chime20b}).

One of the difficulties in studying FRBs has been positional uncertainties of order arcminutes, making follow-up observations at other wavelengths challenging. The discovery of the first FRB to show repeated bursts, FRB\,20121102A \citep{spitler16}, finally enabled radio interferometers to refine the position of an FRB to better than an arcsecond. Over 80 repeaters are now known, approximately half of which have been localised to host galaxies \citep{ibik24}, and with CHIME/FRB outriggers now operational \citep[][and references therein]{2025ApJ...989L..48C}, this number will grow significantly in the coming years.

To further study the nature of FRBs, it is highly desirable to observe the bursts at other wavelengths than radio and hence determine their spectral energy distribution, duration and lags with respect to the radio bursts. This information can then be compared to theoretical models of FRB emission. For example, \citet{metzger19} invoked synchrotron maser emission from ultra-relativistic magnetised shocks, such as the flares produced by ejecta from young magnetars, to explain the millisecond radio bursts in the repeating FRBs, and their model also predicts the presence of brief (millisecond to second) {\em optical} bursts.

Assuming that any optical bursts are approximately coincident with the radio bursts and are of a similar duration (we know this is indeed the case with pulsars, e.g. \citealt{shearer03}), the best way of detecting optical bursts would be to reduce the contribution of the sky as much as possible by taking a continuous sequence of extremely short exposures on a large-aperture telescope in order to catch a radio burst in 1--2 optical frames. Using simultaneous radio observations will tell us precisely which optical frames to co-add in order to assemble the deepest possible image coincident with the radio bursts. Knowing the times of the radio bursts will also allow us to search for any optical emission that might precede or follow the radio bursts, as hypothesised by \citet{metzger19}. This technique was used in an attempt to detect optical pulses from the rotating radio transients (RRATs; \citealt{dhillon11}), and was subsequently applied to FRBs by, for example, \citet{hardy17}, 
% \citet{pilia20} - They did not observe any simultaneous radio bursts of FRB 180916!
% \citet{kilpatrick21} - They used 30.3s exposure times, so not high speed!
\citet{niino22} and \citet{kilpatrick24}. Unfortunately, extragalactic FRBs remain undetected at optical wavelengths, or from any other wavelength or messenger other than radio (e.g. \citealt{zhang24}).

In order to maximise the sensitivity of searches for optical bursts, it is necessary to observe with high-speed cameras mounted on the largest-aperture telescopes, and to target the nearest, and hence brightest, FRBs. To this end, we observed the closest known repeater, FRB\,20200120E \citep{bhardwaj21}, simultaneously with HiPERCAM on the 10.4\,m Gran Telescopio Canarias (GTC; the world's largest optical telescope) and the Effelsberg 100\,m Radio Telescope (the world's largest steerable dish). The results of our study are presented in this paper.

\section{Observations and data reduction}

\subsection{Optical}

\begin{figure}
	\centerline{\includegraphics[width=1.0\columnwidth]{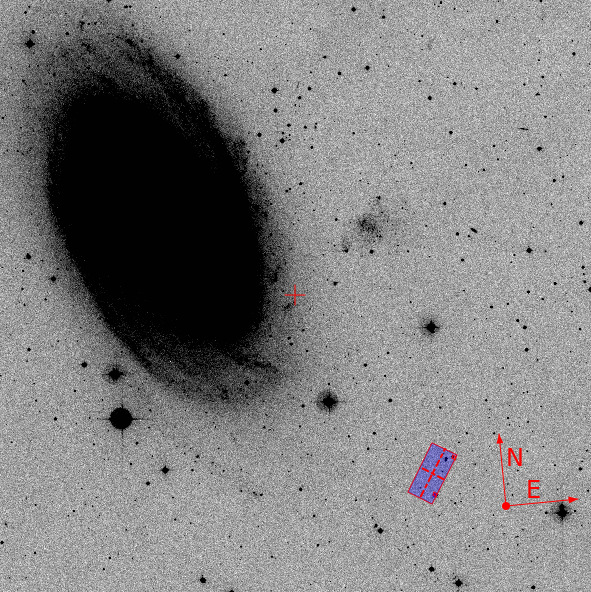}}
    \centerline{\includegraphics[width=\columnwidth]{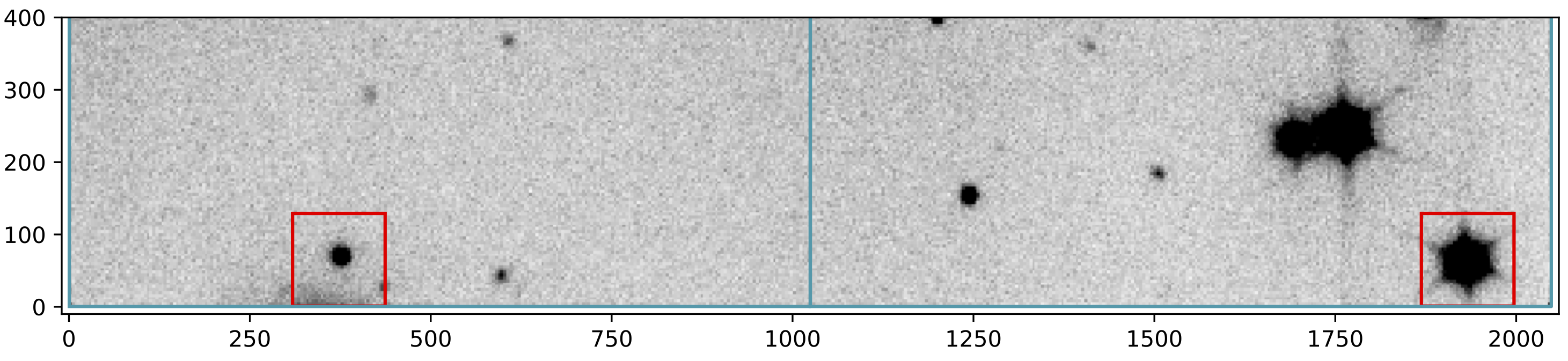}}
    \caption{Top: HiPERCAM finding chart for FRB\,20200120E. The blue rectangle represents the HiPERCAM detector focal plane, which is of diagonal size 3.1$^{\prime}$ and located on the M81 globular cluster containing the FRB. Bottom: Lower portion of the HiPERCAM acquisition image showing the location of the drift-mode windows in red, the right-hand one containing the comparison star and the left-hand one containing the globular cluster --  the centre of the FRB localisation is only 0.116$^{\prime\prime}$ (1.4 unbinned pixels) offset from the optical centre of light of the globular cluster \citep{kirsten22}.}
    \label{fig:fc}
\end{figure}

We observed FRB\,20200120E on the night starting 2024 April 30 with the high-speed, quintuple-beam ($u_s g_s r_s i_s z_s$) optical imager HiPERCAM \citep{dhillon21} mounted at the Folded Cassegrain G focus of the GTC. HiPERCAM was used in drift mode, which gives the highest frame rate (see \citealt{dhillon07}), with one window centred on the position of the M81 globular cluster [PR95]\,30244 that hosts FRB\,20200120E \citep{kirsten22} and the other on a nearby comparison star (SDSS\,J095807.00+685046.8; \citealt{sdss25}), as shown in Fig.~\ref{fig:fc}. The focal-plane mask was used to prevent light from other stars and the sky from falling on regions of the CCD chip above the user-defined windows, which would add noise to the data. Each window was of 128$\times$128 pixels, corresponding to 10.4" on a side, and read out using 4$\times$4 binning, fast parallel clocking and the slow readout speed. A total of 1\,361\,966 frames were obtained between 20:54:04 to 02:03:39 UTC on this night, each of 11.98\,ms exposure time and 1.66\,ms dead time, where each HiPERCAM frame is GPS time-stamped to a relative (i.e. frame-to-frame) accuracy of 0.1\,$\mu$s and an absolute accuracy of 0.1\,ms \citep{dhillon21}. The data were obtained in photometric conditions, with no Moon and seeing of approximately 1".

The data were reduced using the HiPERCAM data reduction pipeline \citep{dhillon21}\footnote{\url{https://hipercam.github.io}}
%https://github.com/HiPERCAM}}. 
All frames were debiased and then flat-fielded, with the latter using the median of twilight-sky frames obtained with the telescope spiralling. The CCD fringing pattern was removed from the $z_s$ HiPERCAM frames using the median of night-sky frames taken with the telescope spiralling. Standard aperture photometry was used to extract counts from [PR95]\,30244 and the comparison star, with object apertures fixed to a radius of 18 unbinned pixels (1.5") and the sky level determined from a clipped mean of the counts in an annulus surrounding each star and subtracted from the object counts. The aperture position of [PR95]\,30244 relative to the comparison star was determined from the sum of $\sim$500\,s of acquisition images, and this offset was then held fixed during the reduction so as to avoid aperture centroiding problems due to the faintness of [PR95]\,30244 in the individual 11.98\,ms science frames. In the $u_s$-band, however, the comparison star was too faint in the individual science frames to act as a reference for the aperture centroiding, so the apertures were held fixed during the reduction.

\subsection{Radio}
\label{sec:radio_obs}

Radio observations were made with the Ultra BroadBand (UBB) receiver of the Effelsberg 100\,m Radio Telescope in Bonn. The receiver covers a frequency range of 1.3$-$6\,GHz, with a gap between 2.6$-$3\,GHz, and an average system equivalent flux density of $\sim$13\,Jy across the band \citep{Limaye26}. The Effelsberg Direct Digitisation (EDD) backend \citep{EDD} sampled and channelised the full receiver band into a total of five sub-bands: two and three adjacent bands below and above the receiver gap, respectively. All five bands were recorded with full polarisation information, and the dispersive sweep within each frequency channel was removed by coherently dedispersing at the source's dispersion measure 
(DM =  87.7527\,\DMunit, \citealt{nimmo22}). The sample time, central frequency, bandwidth and number of frequency channels in each band varied (see Tab.~\ref{tab:ubb_edd} for details). 

Each of the five bands were searched independently for bursts using the \texttt{BLISS}\footnote{\url{https://gitlab1.mpifr-bonn.mpg.de/eff-sp-searching/bliss}} pipeline, which is based on the \texttt{TransientX} single pulse software package \citep{TransientX}. First, channels frequently contaminated with radio frequency interference (RFI) are identified with \texttt{RFIweather}\footnote{\url{https://gitlab1.mpifr-bonn.mpg.de/eff-sp-searching/rfiweather}}. Then, \texttt{filtool}, which is part of \texttt{TransientX}, masked these channels and writes out a new filterbank file. The \texttt{TransientX} search software was then run on these cleaned files, first applying several additional mitigation algorithms (\texttt{kadaneF} and \texttt{zdot}, see \citet{2023ascl} for details). A range of DM trials from 0 to 188\,\DMunit\ were used to 
correct for the remaining dispersion between frequency channels, with the number of trials adapted to the central frequency and bandwidth of each sub-band. These dedispersed time series were boxcar match filtered, with widths ranging from the native time resolution to 3\,ms, in order to optimise the time resolution of the data to the intrinsic width of the burst. A single burst was found in the independent searches of sub-bands 1 and 2 above a signal-to-noise threshold of 6 (see Sect.~\ref{sec:radburst}). No bursts above threshold were found in sub-bands 3 to 5, and all further analysis uses only sub-bands 1 and 2. The \texttt{YOUR} \citep{YOUR} package was used to extract snapshots in the \texttt{hdf5} format from the sub-band 1 and 2 cleaned filterbank files. 

\begin{table}
    \centering
    \begin{tabular}{cccccccc}
     \hline
        Band & $\nu_c$ &  $\Delta \nu_{\rm BW}$ & $\Delta \nu_c$ & $N_{\nu}$ & $t_{\rm samp}$\\
         &  (MHz) &  (MHz) & (MHz) & & ($\mu$s)\\
         \hline
         1  & 1614.843750   & 650 & 0.5078125 & 1280 & 31.51 \\
         2  & 2264.843750  & 650 & 0.5078125 & 1280 & 31.51 \\
         3  & 3539.062500  & 1125 & 0.9375 & 1200 & 32.0 \\
         4  & 4664.062500  & 1125 & 0.9375 & 1200 & 32.0 \\
         5  & 5601.562500  & 750 & 0.9375 & 800 & 32.0 \\
         \hline
    \end{tabular}
    \caption{Properties of the five UBB sub-bands. From left-to-right, the columns show the sub-band label, central frequency, bandwidth, frequency resolution, number of frequency channels, and the time resolution.}
    \label{tab:ubb_edd}
\end{table}

\section{Results}

\subsection{The radio burst}
\label{sec:radburst}

\begin{figure}
    \centerline{\includegraphics[width=\columnwidth]{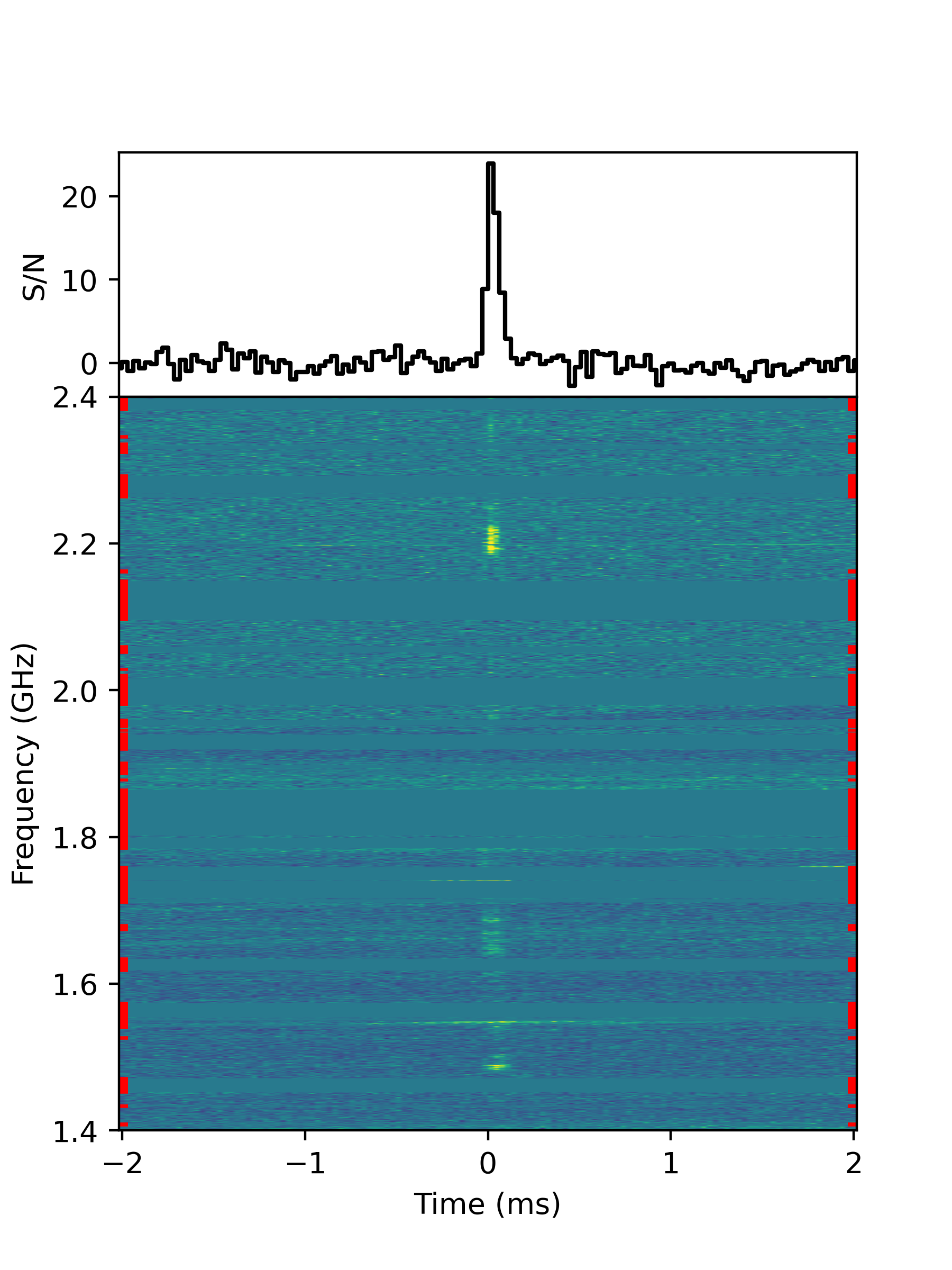}}
    \caption{Bottom: Dynamic spectrum of the radio burst from \frb\ observed simultaneously with HiPERCAM across the lower two UBB sub-bands. The burst was dedispersed with a DM of 87.7527 \DMunit. The time and frequency resolutions are the native resolutions of the data. The red dashes on the plot edges indicate channels removed due to RFI.  The maximum range of the colour scale was set to S/N $= 5$. Top: The profile of the radio burst in units of S/N including only the non-masked channels.}
    \label{fig:frb}
\end{figure}

During the simultaneous observations with HiPERCAM, Effelsberg detected one radio burst 
from \frb. The dynamic spectrum and frequency-averaged profile of the combined sub-band 1 and 2 detections are shown in Fig.~\ref{fig:frb}. The topocentric arrival times of the two detections were 2024 April 30 at 21:52:03.823 UTC (sub-band 2) and 21:52:03.865 UTC (sub-band 1), which corresponds to the expected arrival time difference of 42.5\,ms between the highest frequencies of the two sub-bands due to dispersion. 
%the two adjacent receiver bands centred at 2265\,MHz and 1615\,MHz, respectively, as shown in Fig.~\ref{fig:frb}. 
The radio burst times corrected to infinite frequency and the Solar System barycentre, \texttt{TEMPO} and the DE421 ephemeris, are BMJD (TDB) 60430.9112856426509 and 60430.9112856421052, respectively. The difference between these times of arrival is $\sim$1.5 time samples (i.e.\ 47.1\,$\mu$s), which is less than the measured full-width half maximum of the burst and much less than the 11.98\,ms optical window. We therefore quote a conservative uncertainty based on the observed burst's width (see below) of $\sim$80\,$\mu$s. 

%The burst had a flux density of 0.65$\pm$?\,Jy, signal-to-noise ratio (SNR) 15.7, and width 0.1$\pm$?\,ms, giving a fluence of 0.065$\pm$? Jy\,ms. 
The fluence and observed burst width are calculated after summing the signal across sub-bands 1 and 2 using the extracted snapshots (see Sect.~\ref{sec:radio_obs}). First, a time series in signal-to-noise ratio (S/N) units was calculated excluding the RFI-contaminated channels (Fig.~\ref{fig:frb}, top panel) and then converted to fluence units using the radiometer equation with an assumed system-equivalent flux density of 13\,Jy, the sample time of 31.51\,$\mu$s, a clean bandwidth of 632.2\,MHz, and two summed polarisations. 
The on-pulse samples, defined as those with S/N\,$>$\,2, were then summed,
%Then the on-pulse samples with S/N\,$>$\,2 were summed, 
yielding a final, integrated radio fluence of $\mathcal{F}_{\rm R}=0.14\pm0.01$\,Jy\,ms. The quoted error is simply the statistical error corresponding to the root-mean-square of the time series; systematic uncertainties due to e.g. variations in the telescope sensitivity or RFI are not quantified but are typically around 10\%. The measured full-width-half-maximum burst width, assuming a Gaussian profile, is 79.2$\pm$3.5\,$\mu$s. Our fluence compares with the average burst fluence of 0.17$\pm$0.03\,Jy\,ms from the 60 bursts of FRB\,20200120E observed by \citet{nimmo23}, implying that we observed a fairly typical radio burst.
%-- 29 of the bursts observed by \citet{nimmo23} were fainter than the one we observed.

The isotropic-equivalent (i.e. calculated under the assumption that the radiation is emitted equally in all directions), bandwidth-integrated radio burst energy is determined using the 
equation:
% (e.g. Eqn.~B6 of \citealt{hiramatsu23})
% This formula includes redshift as it takes time dilation into account (which makes the observed pulse width - used in the
% calculation of the fluence - longer than the rest-frame pulse width by a factor 1+z).
\begin{equation}
E_{\rm R} = 9.5\times10^{31} \left(\frac{4\pi}{1+z}\right) \left(\frac{d_{\rm L}}{\rm Mpc}\right)^{2} \left(\frac{\mathcal{F}_{\rm R} \cdot \Delta\nu}{{\rm Jy\,ms \cdot GHz}}\right)\,\,{\rm erg},
\label{eqn:energy}
\end{equation}
where $z$ is the redshift (negligible in this case), $d_{\rm L}$ the luminosity distance (3.63\,Mpc; \citealt{kirsten22}), $\mathcal{F}_{\rm R}$ the radio fluence, and $\Delta\nu$ the frequency bandwidth of the observation. We find a value of $E_{\rm R} = 1.4\times10^{33}$ erg.
% calculation of constant: 1e-26*((1e6*3.086e18)**2)*1e9
% calcualtion of energy: 1e-26*((1e6*3.086e18)**2)*1e9*4*m.pi*(3.63**2.)*0.14*0.6327
% 632.7 MHz assumed for the bandwidth

\subsection{A search for optical bursts}
\label{sec:optburst}

\begin{figure}
    \centerline{\includegraphics[width=\columnwidth]{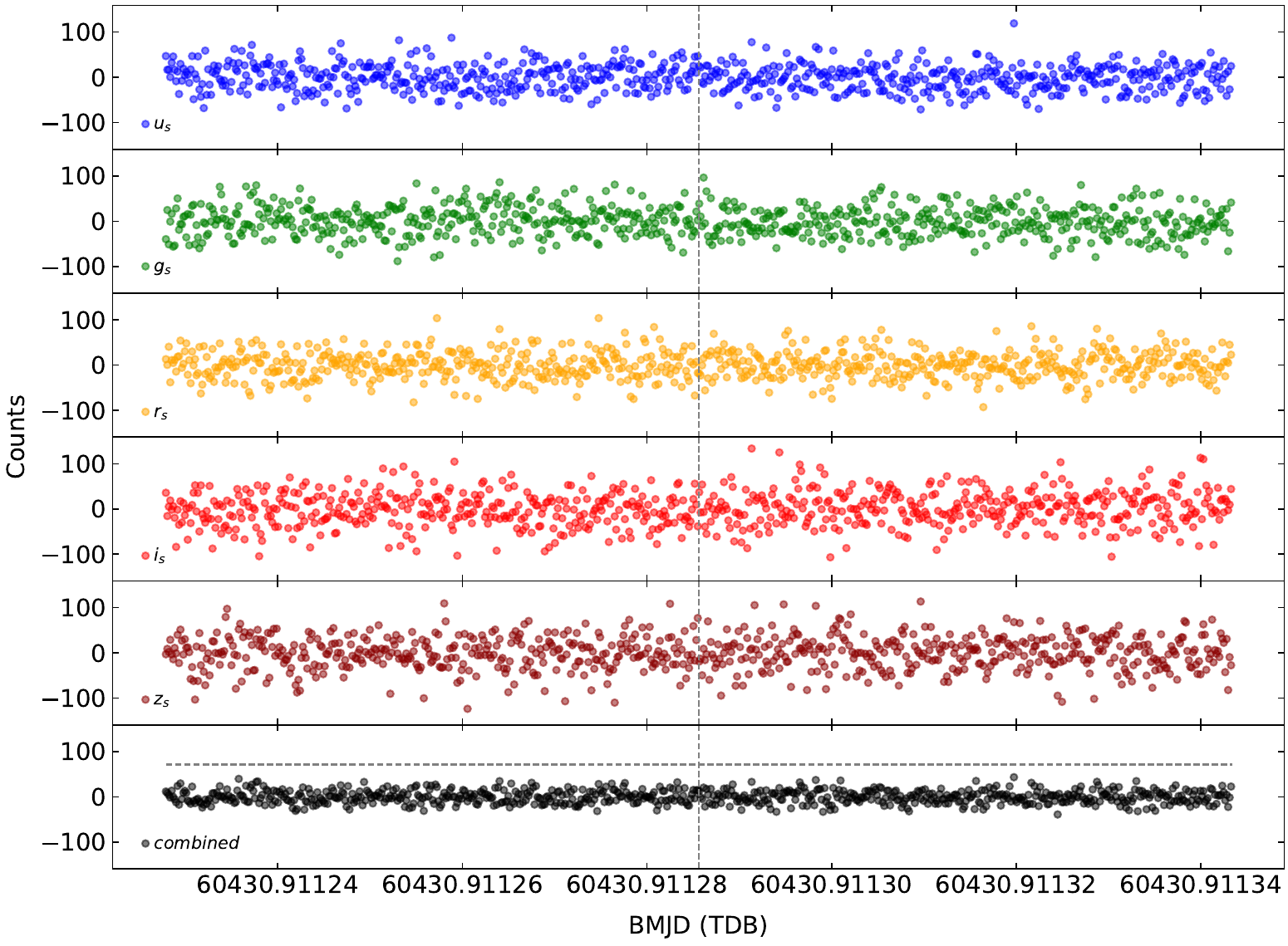}}
    \caption{From top-to-bottom, the HiPERCAM $u_sg_sr_si_sz_s$ and combined light curves of FRB\,20200120E, cropped to 10\,s (733 data points at 13.64\,ms time resolution) around the time of the radio burst detected by Effelsberg, indicated by the vertical dashed black line. These are the sky-subtracted counts in the FRB aperture, with no comparison star correction, and with the mean level due to the globular cluster flux subtracted. The horizontal dashed line in the bottom panel shows the $5\sigma$ level used to derive the limits given in the bottom row of Tab.\,\ref{tab:fluxes}.}
    \label{fig:lc}
\end{figure}

The HiPERCAM light curves of FRB\,20200120E obtained simultaneously with Effelsberg are shown in Fig.~\ref{fig:lc}, cropped to the time around the radio burst detected by Effelsberg. It can be seen that there are no significant outlying HiPERCAM data points coincident with the burst, including in the bottom panel, which shows the combined $u_sg_sr_si_sz_s$ light curve (constructed by taking the inverse-variance weighted mean of the fluxes in each filter, $f_{\text{combined}} = \sum_{j} f_j w_j/\sum_{j} w_j$, where $w_j = 1/\sigma_j^2$ and $j$ is the filter). The radio burst occurred in HiPERCAM frame number 255\,091 of 1\,361\,966, which had start/end times of BMJD (TDB) 60430.9112855833/60430.9112857219,  respectively, and hence the radio burst at BMJD (TDB) 60430.9112856421 occurred almost exactly at the exposure centre and any simultaneous optical burst would not have been missed due to detector readout.

We searched the entire optical light curve for outlying points and found none that could not be explained by cosmic rays falling within the aperture centred on the FRB. This is one of the advantages of having simultaneous observations in multiple optical cameras, as it makes it relatively straightforward to eliminate cosmic rays, which can sometimes produce quite stellar-like point-spread functions and are a key contaminant in single-band studies. We also searched for longer-timescale optical counterparts to the radio burst by binning the data on a range of timescales up to 60\,s and plotting the log of the standard deviation of each binned light curve versus the log of the bin width, finding no significant deviation from the line of gradient $-0.5$ expected for white noise (e.g. \citealt{carter09}). The absence of even longer timescale variability was verified from the significance of fitting a constant to the light curves. In addition, we searched the light curves for periodicities using the Lomb-Scargle algorithm (see \citet{vanderplas18} and references therein), covering frequencies ranging from approximately Nyquist to the inverse of the duration of the observing run. We judged any prominent peaks we found in the periodograms to be insignificant if they appeared in only one filter and/or if they appeared in the periodogram of the comparison star light curve. We found no evidence for any significant periodic signals in our data.

We can place limits on the optical fluence of FRB\,20200120E from our observations by estimating the FRB flux required to be detectable at 5$\sigma$ above the noise in the aperture, where the noise is a combination of the detector noise and shot noise from the sky and globular cluster. We estimated the noise in the FRB aperture from the standard deviation of the 1000 data points centred on the time of radio burst, so as to avoid the effect of any long-term flux variations that may be present in the full light curve. The standard deviation was multiplied by five to give the 5$\sigma$ limit, and then flux calibrated using the known magnitude of the comparison star in each band. This also corrected for atmospheric extinction, which affects both the FRB and comparison star equally, to first order. We then corrected for interstellar extinction using the extinction estimates of \citet{schlafly11}\footnote{\url{https://irsa.ipac.caltech.edu/applications/DUST}}, which assumes a \citet{fitzpatrick99} Galactic extinction law with $R_v = 3.1$ and reddening $E(B-V) = 0.0653$. Finally, the time-integrated (as opposed to bandwidth-integrated) fluence limits were calculated by multiplying the flux limits by the 11.98\,ms exposure time. The resulting limits on the optical burst fluence and optical-to-radio fluence ratios of FRB\,20200120E, together with the various input values used in the calculations, are given in Tab.~\ref{tab:fluxes}. The bottom row of this table gives the limits for the combined light curve, calculated using $f_{5\sigma, \text{combined}} = (\sum_{j} 1 / f_{5\sigma, j}^2)^{-1/2}$, where $f_{5\sigma, j}$ are the calibrated flux limits for each filter $j$ given in the fifth column of Tab.~\ref{tab:fluxes}, and we have effectively assumed a flat spectrum source. The combined limits provide the deepest values and hence whenever we quote 
an optical fluence limit for FRB\,20200120E in this paper, we are referring to the combined $u_s g_s r_s i_s z_s$ band. Note that the shot noise due to light from [PR95]\,30244 (SDSS\,J095754.71+684900.7), which is of magnitude $u$/$g$/$r$/$i$/$z$ = 21.69/20.20/19.56/19.36/19.17 \citep{sdss25}, is negligible compared to the $\sim$4\,e$^-$/pixel readout noise of the HiPERCAM CCDs in the 11.98\,ms exposure time, and hence we can ignore the impact of shot noise from the globular cluster in the preceding analysis. Given that we detected no points lying above 5$\sigma$ in the entire light curve, the limits given in Tab.~\ref{tab:fluxes} are applicable not only to optical emission simultaneous with the radio burst, but also to any precursor or delayed optical emission of duration $\leq$\,11.98\,ms during the $\sim$\,5\,hr observing window.

\begin{table}
 \caption{Limits on the optical burst fluence and optical-to-radio fluence ratio of FRB\,20200120E in each filter, and in the combination (inverse-variance weighted mean) of all five filters, tabulated together with the values involved in the calculations: comparison star magnitude, $m_{\rm comp}$; interstellar extinction, $A$; 5$\sigma$ lower limit on the optical magnitude of FRB\,20200120E in the 11.98\,ms exposure time, $m_{\rm 5\sigma}$; 5$\sigma$ upper limit on the optical flux of FRB\,20200120E in the 11.98\,ms exposure time, $f_{\rm 5\sigma}$; 5$\sigma$ upper limit on the optical fluence, $\mathcal{F}_{\rm O}$; upper limit on the optical-to-radio fluence ratio, 
 $\mathcal{F}_{\rm O}/\mathcal{F}_{\rm R}$. Note that all of the limits presented in this table have been corrected for interstellar extinction.
 }
 \label{tab:fluxes}
\begin{center}
 \begin{tabular}{ccccccc}
  \hline
Filter & $m_{\rm comp}$ & $A$ & $m_{\rm 5\sigma}$ & $f_{\rm 5\sigma}$ & $\mathcal{F}_{\rm O}$ & $\mathcal{F}_{\rm O}/\mathcal{F}_{\rm R}$ \\
& (mag) & (mag) & (mag) & (mJy) & (Jy\,ms) & \\
  \hline
$u_s$    & 17.41 & 0.32 & $>$16.4 & $<$0.98 & $<$0.0117  & $<$0.084 \\
$g_s$    & 15.69 & 0.25 & $>$18.0 & $<$0.22 & $<$0.0027  & $<$0.019 \\
$r_s$    & 15.05 & 0.17 & $>$17.8 & $<$0.26 & $<$0.0032  & $<$0.023 \\
$i_s$    & 14.82 & 0.13 & $>$17.4 & $<$0.41 & $<$0.0049  & $<$0.035 \\
$z_s$    & 14.72 & 0.09 & $>$17.0 & $<$0.55 & $<$0.0066  & $<$0.047 \\
%combined &       &      &         & $<$0.13 & $<$0.0015  & $<$0.011 \\
combined &       &      &         & $<$0.15 & $<$0.0018  & $<$0.013 \\
  \hline
 \end{tabular}
\end{center}
\end{table}

The isotropic-equivalent, bandwidth-integrated optical burst energy is defined by Eqn.~\ref{eqn:energy} with $\mathcal{F}_{\rm R}$ replaced by the $\mathcal{F}_{\rm O}$ value for $r_s$ given in Tab.~\ref{tab:fluxes} (which is actually slightly more constraining than the combined data due to the narrower bandpass) and using $\Delta\nu_{r_{s}} = 107\,192$\,GHz. We find that the optical energy of the FRB\,20200120E burst must be $<$\,$5.4\times$10$^{36}$\,erg, otherwise we would have detected it in our $r_s$-band data. 

\section{Discussion}

\begin{figure*}
    \centerline{\includegraphics[width=2\columnwidth]{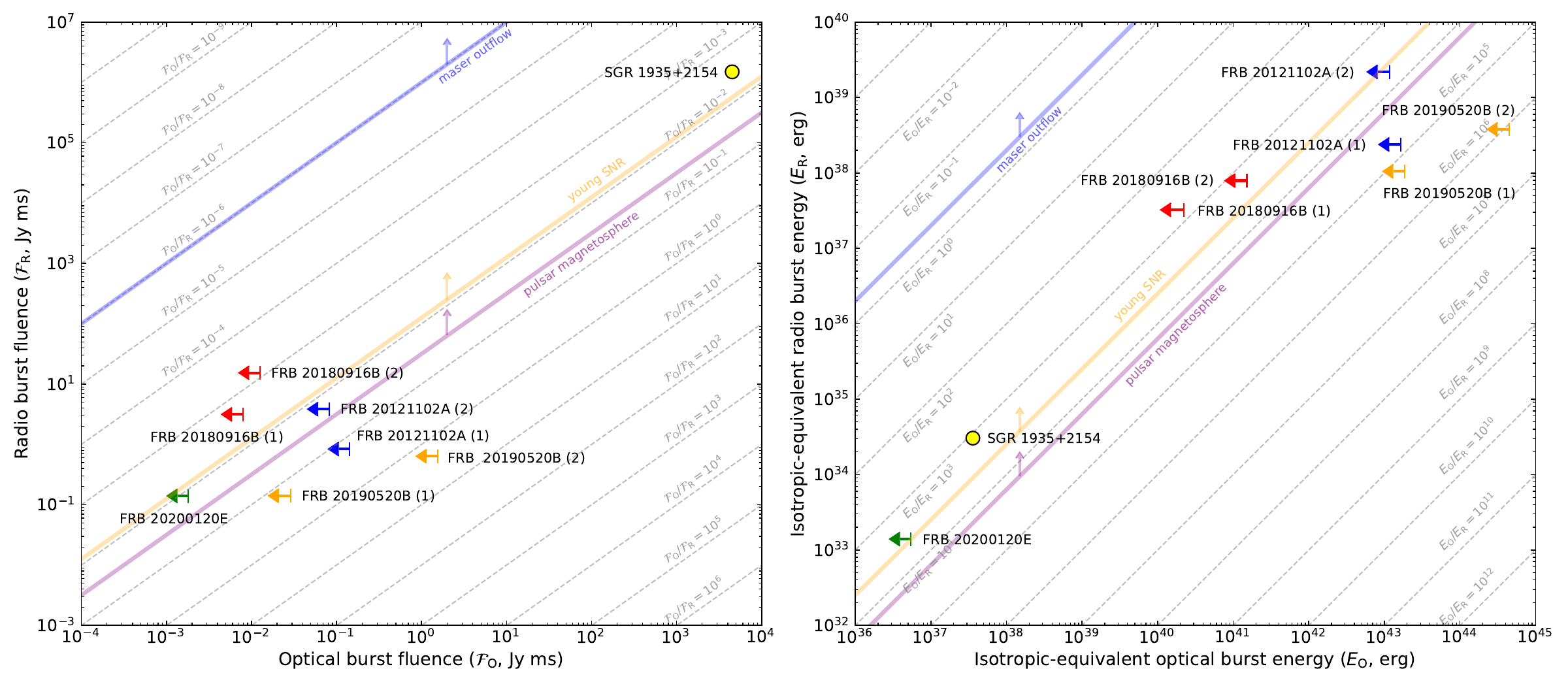}}
    \caption{Left: Radio burst fluence of FRB\,20200120E versus the upper limit on the optical fluence (5$\sigma$, combined data, green arrow). For comparison, we also plot the upper limits for FRB\,20180916B (red arrows; $1 = $ \citet{kilpatrick24}, 3$\sigma$, $r$-band; $2 = $  \citet{trudu23}, 3$\sigma$, 320--900\,nm), FRB\,20190520B (orange arrows; $1 = $ \citet{niino22}, 5$\sigma$, 370-730\,nm; $2 = $ \citet{yan25}, 3$\sigma$, $U$-band), FRB\,20121102A (blue arrows; $1 = $ \citet{hardy17}, 5$\sigma$, $i$+$z$-band; $2 = $ \citet{magic18}, 5$\sigma$, $U$-band), and an interpolated value for the Galactic magnetar SGR\,1935+2154 (yellow circle; radio: \citealt{bochenek20}; optical ($i$-band): \citealt{kilpatrick24}). All values have been corrected for interstellar extinction. The grey dashed lines show the optical-to-radio fluence ratio, $\mathcal{F}_{\rm O}/\mathcal{F}_{\rm R}$. The coloured lines show the predicted $\mathcal{F}_{\rm O}/\mathcal{F}_{\rm R}$ upper limits for inverse Compton scattering in a pulsar magnetosphere (purple), young supernova remnant (orange), and maser outflow (blue) -- taken from Fig.\,3 of \citet{yang19}. Right: Isotropic-equivalent, bandwidth-integrated radio burst energy of FRB\,20200120E versus the upper limit on the optical burst energy (5$\sigma$, $r_s$-band, green arrow). The grey dashed lines show the optical-to-radio energy ratio, ${E}_{\rm O}/{E}_{\rm R}$. The points in this panel are calculated from the fluences plotted in the left-hand panel, and the coloured points and lines have the same meanings.}
    \label{fig:limits}
\end{figure*}

Fig.~\ref{fig:limits} (left) shows the radio burst fluence $\mathcal{F}_{\rm R}$ of FRB\,20200120E versus the upper limit on the optical burst fluence $\mathcal{F}_{\rm O}$ derived in this paper, compared with those of other FRBs that have been subjected to optical searches with high-speed cameras: FRB\,20180916B (\citealt{kilpatrick24}, \citealt{trudu23}), FRB\,20190520B (\citealt{niino22}, \citealt{yan25}) and FRB\,20121102A (\citealt{hardy17}, \citealt{magic18}). We also plot an interpolated value for the Galactic magnetar SGR\,1935+2154 (see below). All of the values plotted in 
Fig.~\ref{fig:limits} have been corrected for interstellar extinction. In terms of optical fluence, the limit we have placed on FRB\,20200120E is the lowest to date for an FRB, which is unsurprising given that we used a high-speed, quintuple-beam camera mounted on the largest-aperture optical telescope. However, due to the faintness of the radio burst we detected from FRB\,20200120E with Effelsberg, the upper limit on the optical-to-radio fluence ratio, $\mathcal{F}_{\rm O}/\mathcal{F}_{\rm R}$, is most constrained by the observations of FRB\,20180916B by \citet{trudu23}, noting that this ratio {\em decreases} as one moves up the $y$ axis of Fig.~\ref{fig:limits} (left).

To place these results in context, we show the predicted limits on $\mathcal{F}_{\rm O}/\mathcal{F}_{\rm R}$ for various models of FRB optical bursts from \citet{yang19} involving inverse-Compton scattering from a highly-magnetised central source, like a magnetar or young pulsar. Whereas $\mathcal{F}_{\rm O}$ constrains the absolute energetics of the source, which can be affected by factors such as the size and temperature of the optically-emitting region, the ratio $\mathcal{F}_{\rm O}/\mathcal{F}_{\rm R}$ tests the underlying radiative processes, such as whether they are coherent or incoherent. Three such models are plotted in Fig.~\ref{fig:limits} -- emission from a pulsar magnetosphere, pulsar nebula (young supernova remnant; SNR\footnote{It should be noted that this model is unlikely to apply to FRB\,20200120E given its location in a globular cluster.}) and maser outflow. Note that we do not consider here the other models explored by \citet{yang19} -- curvature radiation and direct maser emission (as opposed to the inverse-Compton scattering from the maser electrons considered above) -- since these predict optical counterparts with much lower $\mathcal{F}_{\rm O}/\mathcal{F}_{\rm R}$ values than plotted in Fig.~\ref{fig:limits}. We also do not consider here the synchrotron maser emission model of \citet{metzger19}, which predicts much longer timescale (up to hours) optical afterglows from relativistic plasma ejected from a magnetar that creates a shock in the external magnetized medium (produced in previous flare events) -- see also Appendix~B of \citet{hiramatsu23}. The coloured diagonal lines in Fig.~\ref{fig:limits} indicate the {\em maximum} value of the $\mathcal{F}_{\rm O}/\mathcal{F}_{\rm R}$ ratio predicted by each model, and hence any data points lying diagonally above a line are compatible with that model. It can be seen that our optical-to-radio fluence ratio for FRB\,20200120E is within the model predictions for a pulsar magnetosphere and close to compatibility with the pulsar nebula model. The same is true for both FRB\,20180916B\footnote{Here, we disagree with the conclusion of \citet{kilpatrick24}, as $\mathcal{F}_{\rm O}/\mathcal{F}_{\rm R}$ for this object falls just below the upper bounds of both models and is hence compatible with them.} and SGR\,1935+2154. It is important to note, however, that the environments of the four extragalactic FRBs plotted in Fig.~\ref{fig:limits} are quite different, and hence their progenitors and emission mechanisms may also be quite different: FRB\,20200120E is in a globular cluster, FRB\,20180916B is in a star-forming region in a massive spiral galaxy \citep{marcote20}, and FRB\,20121102A and FRB\,20190520B are in star-forming, low-metallicity dwarf galaxies (\citealt{bassa17}, \citealt{chen25}).

The optical limits presented in this paper are arguably most constraining when it comes to considering the burst energy rather than fluence, as FRB\,20200120E is approximately 40 times closer than the next closest extragalactic FRB plotted in Fig.~\ref{fig:limits} (FRB\,20180916B at 150\,Mpc, \citealt{kilpatrick24}). Hence in Fig.~\ref{fig:limits} (right) we plot the isotropic-equivalent radio burst energy of FRBs versus their optical burst energy. The optical ($r_s$-band) limit we derived for FRB\,20200120E is almost four orders of magnitude lower in energy than for other extragalactic FRBs, and an order of magnitude lower than for the Galactic magnetar SGR\,1935+2154, although we caution the reader that no optical emission has ever been observed from SGR\,1935+2154 and the values presented in Fig.~\ref{fig:limits} are based on an interpolation of the X-ray and radio detections during the FRB-like burst observed from this object \citep{kilpatrick24}. In fact, no bursts or flares have been detected in the optical for {\em any} soft gamma repeaters (SGRs), and the only quasi-persistent (pulsed) optical emission ever detected from an SGR (SGR\,0501+4516) had an isotropic-equivalent optical energy of only $\sim$2.8$\times$10$^{27}$\,erg \citep{dhillon11b}, far below the limit we have derived here.

The search for prompt X-ray emission simultaneous with radio bursts from FRB\,20200120E by \citet{pearlman25} also revealed no high-energy counterpart, although their X-ray limits were not sensitive enough to detect the FRB-like bursts observed in the X-rays from SGR\,1935+2154 when placed at the distance of FRB\,20200120E. Their observations were sensitive enough, however, to rule out an association between FRB\,20200120E and magnetar giant flares, intermediate magnetar flares, and ultra-luminous X-ray bursts (ULXBs).

%Moreover, their limits of persistent X-ray luminosity rule out PWn similar to the Crab pulsar and a range of ultra-luminous and accreting X-ray sources. The observations were fully consistent with a pulsar that emits giant pulses or a magnetar formed through a delayed channel. 

In the future, our optical limits could be improved upon by observing a highly active repeater during a burst "storm", as has been observed for FRB\,20200120E \citep{nimmo23}, and stacking the optical frames containing the radio bursts together. For example, detecting 50 radio bursts would improve the optical fluence limit by a factor of $\sim$7, assuming the bursts are of similar energies, although we know that FRB\,20200120E does on occasion exhibit radio bursts with fluences nearly 200 times greater than the mean \citep{zhang24b}. Capturing such bursts would require a joint optical/radio target-of-opportunity campaign on currently active repeaters, and a certain amount of luck. 

In general, searches for high-energy, bursting counterparts to FRBs benefit from having nearby sources and bright radio bursts. Both of these goals are accomplished with aperture array telescopes such as BURSTT-256 \citep{BURSTT} and CASM \citep{CASM} that have instantaneous fields of view of $\sim$10$^4$\,deg$^2$ and sub-arcsecond localisation. Such surveys find nearby bursts ($z$\,$\lesssim$\,0.05) with high fluences ($\gtrsim 100$\,Jy\,ms). Dedicated, deep follow-up of the repeaters found by these surveys with high-speed, multi-band optical cameras mounted on the largest-aperture telescopes is likely to provide the best chance of detecting FRBs in the optical.

\section{Conclusions}

We have presented simultaneous optical and radio observations of the repeating FRB\,20200120E, the nearest extragalactic FRB, located in a globular cluster of M81. We detected a single, weak radio burst with a fluence of $\mathcal{F}_{\rm R} = 0.14\pm0.01$\,Jy\,ms. No corresponding optical burst was detected in the simultaneous $u_s g_s r_s i_s z_s$ frames, to a limiting fluence of $\mathcal{F}_{\rm O} < 1.8\times10^{-3}$\,Jy\,ms (combined bands) and energy ${E}_{\rm O}<5.4\times10^{36}$\,erg ($r_s$-band) in the 11.98\,ms exposure. The derived optical-to-radio fluence ratio $\mathcal{F}_{\rm O}/\mathcal{F}_{\rm R} < 0.013$ is compared to various emission models involving inverse-Compton scattering in the magnetosphere or nebula of a highly-magnetised neutron star. 

\section*{Acknowledgements}
We thank the anonymous referee for their careful reading of the manuscript. The construction of HiPERCAM was funded by the European Research Council under the European Union's Seventh Framework Programme (FP/2007-2013) under ERC-2013-ADG Grant Agreement no. 340040 (HiPERCAM). The operation of HiPERCAM is funded by the United Kingdom's Science and Technology Facilities Council under grant ST/Z000033/1. Based on observations made with the Gran Telescopio Canarias, installed at the Spanish Observatorio del Roque de los Muchachos of the Instituto de Astrof\'{i}sica de Canarias, on the island of La Palma. This work also makes use of observations obtained with the 100-m telescope of the Max-Planck-Institut für Radioastronomie (MPIfR) at Effelsberg. The Effelsberg EDD backend system is developed and maintained by the electronics division at MPIfR. VSD is supported by a Leverhulme Research Fellowship (RF-2025-297/9), the Spanish Ministry of Science, Innovation and Universities (PID2023-151588NB-I00), and the Agencia Estatal de Investigación (AEI) through the Severo Ochoa Centre of Excellence accreditation awarded to the Instituto de Astrofísica de Canarias, grant CEX2025-001609-S, funded by MICIU/AEI/10.13039/501100011033. LGS is a Lise Meitner Max Planck Research Group leader.

\section*{Data Availability}
HiPERCAM data are available through the GTC public archive
\footnote{\url{https://gtc.sdc.cab.inta-csic.es/gtc}}.

\bibliographystyle{mnras}
\bibliography{hipercam_frb}
\bsp
\label{lastpage}
\end{document}